\documentclass[twocolumn,amstext,amssymb,superscriptaddress,floatfix,showpacs,nofootinbib]{revtex4-1}
\usepackage{amsmath}
\usepackage{lineno} 
\usepackage{ulem}

\usepackage{todonotes}
\presetkeys{todonotes}{color=green!40, size=\footnotesize}{}
\usepackage{hyperref}

\hypersetup{colorlinks=true,citecolor=blue,citebordercolor=red,linktoc=all,linkcolor=blue}

\usepackage{setspace}

\usepackage{amssymb,amsfonts,multirow}

\usepackage{epsfig}

\usepackage{amssymb,url,bm}
\usepackage{graphicx}
\usepackage{hyperref}

\usepackage{color}

\usepackage{array}
\usepackage{amsmath}
\usepackage{slashed}

\long\def\del #1 \enddel { }

\usepackage{epsfig}

\usepackage{xcolor,colortbl}

\definecolor{Gray}{gray}{0.85}
\definecolor{LightGreen}{rgb}{0.88, 1, 0.88}
\definecolor{Blue}{rgb}{0,1,1}
\definecolor{Lime}{rgb}{0,1,0}
\definecolor{LightCyan}{rgb}{0.88,1,1}
\definecolor{LightRed}{rgb}{1, 0.85, 0.85}
\definecolor{Red}{rgb}{1, 0, 0}
\definecolor{LightYellow}{rgb}{1, 1, 0.85}
\definecolor{Yellow}{rgb}{1,1,0.05}
\definecolor{LightBlue}{rgb}{0.87, 0.94, 1}
\definecolor{white}{gray}{1}
\definecolor{black}{gray}{0}

\definecolor{LightGray}{gray}{0.93}

\usepackage{epsfig}

\usepackage{array,mathtools,amssymb,booktabs}
\newcolumntype{C}{>{$}c<{$}}
\AtBeginDocument{
\heavyrulewidth=.16em
\lightrulewidth=.1em
\cmidrulewidth=.03em
\belowrulesep=.4ex
\belowbottomsep=0pt
\aboverulesep=.4ex
\abovetopsep=0pt
\cmidrulesep=\doublerulesep
\cmidrulekern=.5em
\defaultaddspace=.5em
}

\usepackage{hyperref}

\newcolumntype{G}{>{\columncolor{LightGray}}c}

\usepackage{hyperref}

\def\beq{\begin{equation}}
\def\eeq{\end{equation}}

\def\bea{\arraycolsep .1em \begin{eqnarray}}
\def\eea{\end{eqnarray}}
\def\Tr{{\rm Tr}}

\def\eps{\epsilon}

\def\al#1{\alpha_{{#1}}}

\def\eq#1{(\ref{#1})}

\def\s0#1#2{\mbox{\small{$ \frac{#1}{#2} $}}}
\def\0#1#2{\frac{#1}{#2}}

\def\al#1{\alpha_#1}
\def\grgl{\:\hbox to -0.2pt{\lower2.5pt\hbox{$\sim$}\hss}{\raise3pt\hbox{$>$}}\:}
\def\klgl{\:\hbox to -0.2pt{\lower2.5pt\hbox{$\sim$}\hss}{\raise3pt\hbox{$<$}}\:}

\def\lsim{\mathrel{\rlap{\lower4pt\hbox{\hskip1pt$\sim$}}
    \raise1pt\hbox{$<$}}}                
\def\gsim{\mathrel{\rlap{\lower4pt\hbox{\hskip1pt$\sim$}}
    \raise1pt\hbox{$>$}}}                

\makeatletter
    \def\CT@@do@color{%
      \global\let\CT@do@color\relax
            \@tempdima\wd\z@
            \advance\@tempdima\@tempdimb
            \advance\@tempdima\@tempdimc
    \advance\@tempdimb\tabcolsep
    \advance\@tempdimc\tabcolsep
    \advance\@tempdima2\tabcolsep
            \kern-\@tempdimb
            \leaders\vrule
                    \hskip\@tempdima\@plus  1fill
            \kern-\@tempdimc
            \hskip-\wd\z@ \@plus -1fill }
    \makeatother
\begin{document}
${}$\vskip1cm

\title{UV conformal window for asymptotic safety}
\author{Andrew~D.~Bond}
\email{a.bond@sussex.ac.uk}
\author{Daniel F.~Litim}
\email{d.litim@sussex.ac.uk}
\author{Gustavo Medina Vazquez}
\email{g.medina-vazquez@sussex.ac.uk}
\author{Tom Steudtner}
\email{t.steudtner@sussex.ac.uk}
\affiliation{\mbox{Department of Physics and Astronomy, U Sussex, Brighton, BN1 9QH, U.K.}}

\begin{abstract}
Interacting fixed points in four-dimensional gauge theories coupled to matter are investigated using perturbation theory up to three loop order. It is shown how fixed points, scaling exponents, and anomalous dimensions  are obtained as a systematic power series in a small parameter. The underlying ordering principle is explained and contrasted with  conventional perturbation theory and Weyl consistency conditions. 
We then determine the conformal window with asymptotic safety from the complete next-to-next-to-leading order in perturbation theory.  Limits for the conformal window  arise  due to fixed point mergers, the onset of strong coupling, or vacuum instability.
A consistent picture is uncovered by comparing various levels of approximation. The theory remains perturbative in the entire conformal window, with vacuum stability dictating the tightest constraints. We also speculate about a secondary conformal window at strong coupling and estimate its lower limit. 
Implications for model building and cosmology are indicated.
\end{abstract}
\vskip-2cm

\maketitle

\section{\bf Introduction}

In recent years the asymptotic safety conjecture \cite{Weinberg:1980gg} has grown into a powerful paradigm of its own,  with many applications ranging  from quantum gravitation to particle physics and critical phenomena \cite{Litim:2011cp}.
It states that quantum field theories remain well-defined and predictive up to highest energies
provided they are governed by an interacting UV fixed point under their renormalisation group evolution of couplings  \cite{Wilson:1971bg}. Asymptotic safety generalises the notion of asymptotic freedom \cite{Gross:1973id,Politzer:1973fx}. The most striking new effects are residual interactions in the UV  which modify canonical power counting and the dynamics of theories at shortest distances \cite{Falls:2013bv}. 

Asymptotic safety has originally been proposed to cure the high energy behaviour of $4d$ quantum gravity by means of an interacting UV fixed point \cite{Weinberg:1980gg}. A lot of progress has been made over the past decades to substantiate the feasibility for an asymptotically safe version of quantum gravity
\cite{Reuter:1996cp,Litim:2003vp,Litim:2006dx,Niedermaier:2006ns,
Litim:2008tt,Codello:2008vh,Litim:2011cp,Falls:2013bv,Falls:2014tra}.
In $3d$  settings,  asymptotic safety is known to arise in models with scalars, or fermions, or both. 
In suitable large-$N$ limits, exact results at weak coupling are available
from the renormalisation group \cite{Bardeen:1983rv,Gawedzki:1985ed,deCalan:1991km,Braun:2010tt}, including models with supersymmetry or spontaneously broken scale invariance \cite{Litim:2011bf,Heilmann:2012yf,Marchais:2017jqc}.
Lattice results are available for non-linear sigma models \cite{Wellegehausen:2014noa}.  More recently, it has  been discovered that asymptotic safety is operative in $4d$ gauge theories with matter
\cite{Litim:2014uca}. 
For this to happen at weak coupling, all three types of elementary fields -- gauge fields, fermions, and scalars -- are required, together with suitable Yukawa couplings \cite{Bond:2016dvk}. By now, necessary and sufficient conditions alongside strict no-go theorems  for asymptotic safety of general gauge theories are known \cite{Bond:2016dvk,Bond:2017sem}.  
 Explicit proofs for asymptotic safety have been given for simple \cite{Litim:2014uca}, semi-simple \cite{Bond:2017lnq} and supersymmetric gauge theories coupled to  matter \cite{ Bond:2017suy}. Coleman-Weinberg resummations  \cite{Litim:2015iea}, the impact of interactions with negative canonical mass dimensions \cite{Buyukbese:2017ehm} and fixed points for models away from $4d$ \cite{Codello:2016muj} have also been investigated. Asymptotically safe extensions of the Standard Model and their signatures at colliders have first been put forward in \cite{Bond:2017wut}.

An important open question relates to the size of the conformal window for asymptotically safe gauge theories, meaning the range in parameter space where a viable  interacting UV fixed point persists. While interacting UV  fixed points are under good control at weak coupling, much less is known  about asymptotic safety  at strong coupling \cite{Falls:2013bv}.  
On the other hand, IR conformal windows of QCD-like theories have been studied more extensively. There, conformal windows are known to extend into the domain of strong coupling \cite{Appelquist:2007hu,Dietrich:2006cm,Braun:2009ns,DelDebbio:2010zz}. Similar insights into UV conformal windows  would be most useful, both conceptually, and from the viewpoint of phenomenology and model building. 

In this paper, we access the conformal window with the help of perturbation theory.
It is shown how fixed points, scaling exponents, and anomalous dimensions  are obtained as a systematic power series in a small parameter  (Sect.~\ref{AS}).  We analyse the systematics  of perturbative approximations for general theories with weakly interacting fixed points and compare the  ordering principle with  conventional perturbation theory and Weyl consistency condition. 
The work of \cite{Litim:2014uca} is extended to derive the requisite beta functions, fixed points, anomalous dimensions, and scaling exponents 
at the complete next-to-next-to-leading order  (Sect.~\ref{2NLO}).
A consistent picture for the conformal window is uncovered by comparing various levels of approximation, with vacuum stability offering the tightest constraints   (Sect.~\ref{UV}). Implications for model building and cosmology are indicated as well. We close with a brief discussion (Sect.~\ref{discussion}). Some technicalities are summarised in an Appendix (App.~\ref{Tech}).

\section{\bf Asymptotic safety}\label{AS}
In this section, we recall the model of \cite{Litim:2014uca} in the Veneziano limit, and provide its beta function for all canonically massless couplings up to 3-loop (2-loop) order in the gauge (Yukawa, scalar) beta functions, and all anomalous dimensions up to 2-loop. We also discuss the underlying systematics for expansions in perturbation theory. 

\subsection{The model}
We consider $4d$ massless quantum field theories with $SU(N_C)$ gauge fields $A^a_\mu$ with field strength $F^a_{\mu\nu}$, coupled to $N_F$ flavors of fermions $Q_i$ in the fundamental representation. The theory also contains a scalar singlet ``meson'' field $H$, a $N_F\times N_F$ complex matrix uncharged under the gauge group, which interacts with the fermions via a Yukawa term. The theory has a global $SU(N_F)\times SU(N_F)$ flavor symmetry. 
The  action is taken to be the sum of the Yang-Mills action, the fermion and scalar kinetic terms, the Yukawa term, and the scalar self-interaction Lagrangean 
\beq
\label{L}
L=L_{\rm YM} + L_{\rm kin.}+L_{\rm Yuk.}+L_{\rm pot.}
\eeq
where
\beq\label{Lall}
\begin{array}{rcl}
L_{\rm YM}&=& - \frac{1}{2} \Tr \,F^{\mu \nu} F_{\mu \nu}\\[1ex]
L_{\rm kin.}&=& \Tr\left(
\overline{Q}\,  i\slashed{D}\, Q \right)
+\Tr\,(\partial_\mu H ^\dagger\, \partial^\mu H) \\[1ex]
L_{\rm Yuk.}&=&-y\,\Tr\left( \overline{Q}_L\, H\, Q_R  \right)  +{\rm h.c.}
\\[1ex]
L_{\rm pot.}&=&
-u\,\Tr\,(H ^\dagger H\,H ^\dagger H )
-v\,(\Tr\,H ^\dagger H )^2  \,. 
\end{array}
\eeq
$\Tr$ is the trace over both color and flavor indices, and the decomposition $Q=Q_L+Q_R$ with $Q_{L/R}=\frac 12(1\pm \gamma_5)Q$ is understood. The theory has four canonically marginal couplings given by the gauge coupling $g$, the Yukawa $y$ and two quartic scalar couplings $u$ and $v$. The theory is renormalisable in perturbation theory.

\subsection{Veneziano limit}
To prepare for the Veneziano (large-$N$) limit with finite couplings \cite{Veneziano:1979ec}, 
we rescale the four canonically dimensionless couplings with suitable powers of field multiplicities,
\beq\label{couplings}
\begin{array}{l}
\displaystyle
\al g=\frac{g^2\,N_C}{(4\pi)^2}\,,\quad
\al y=\frac{y^{2}\,N_C}{(4\pi)^2}\,,\\[3ex]
\displaystyle
\al u=\frac{{u}\,N_F}{(4\pi)^2}\,,\quad \,
\al v=\frac{{v}\,N^2_F}{(4\pi)^2}\,.
\end{array}
\eeq
The theory is then characterised by two free parameters  $N_C$ and $N_F$, related to the field multiplicities. In the Veneziano limit, these are send to infinity while the ratio is kept fixed. This procedure reduces the set of free parameters down to one, which we chose to be
\begin{equation}\label{eps}
\eps=\frac{N_F}{N_C}-\frac{11}{2}\,.
\end{equation}
In the Veneziano limit, $\eps$ is a continuous parameter taking values within $[-11/2,\infty]$. For $\eps<0$, the theory is asymptotically free in all couplings. Trajectories running out of the Gaussian fixed point are trivially ``UV complete''. For $\eps>0$, asymptotic freedom of the gauge sector is lost. In this regime, and for sufficiently small $\eps$, the theory develops an interacting UV fixed point.  Strict perturbative control for an asymptotically safe UV fixed point  is guaranteed as long as
\beq\label{small}
0\le\eps\ll 1\,,
\eeq 
which is the regime of interest for the rest of this work.

\subsection{Renormalisation group}\label{RG}
Quantum effects and the energy-dependence of couplings are encoded in the RG beta functions, which are obtained in the $\overline{\mathrm{MS}}$ renormalisation scheme 
\cite{Machacek:1983tz,Machacek:1983fi,Machacek:1984zw,Luo:2002ti,Pickering:2001aq}. For small coupling, the perturbative loop expansion is reliable, and we write
\beq\label{betaPT}
\beta=\beta^{(1)}+\beta^{(2)}+\beta^{(3)}+\cdots
\eeq
for any of the beta functions $\beta\equiv d\alpha/d\ln\mu$. Here, we denote with $\beta^{(n)}$ the $n^{\rm th}$ loop contribution. Some technicalities in the derivation of beta functions from general expressions 
are summarised in App.~\ref{Tech}.

In concrete terms, the gauge beta function $\beta_g$   up to three loops is given by
  \beq\label{betag}
\begin{array}{rcl}
\beta^{(1)}_g&=&
\displaystyle
\0{4}{3}\eps \,\alpha_g^2 \,,
\\[2ex]
\beta^{(2)}_g&=&
\left(25+\0{26}{3}\eps\right) \alpha_g^3
-2\left(\0{11}{2}+\eps\right)^2\,\alpha_y 
\,\alpha_g^2 \,,
\\[2ex]
\beta^{(3)}_g&=&
\left(\0{701}{6}+  \0{53}{3} \eps - \0{112}{27} \eps^2\right) \alpha_g^4 
\\[2ex] && 
-    \0{27}{8} (11 + 2 \eps)^2 \alpha_g^3 \alpha_y 
\\[2ex] && 
+ \0{1}{4} (11 + 2 \eps)^2 (20 + 3 \eps) \alpha_y^2\alpha_g^2\,.
   \end{array}
\eeq
Up to three loop, the running of the gauge coupling is only sensitive to the gauge and Yukawa coupling. Subleading terms of the order $\sim 1/N_F$ and $\sim 1/N_C$ do not contribute in the Veneziano limit and have been suppressed. 

The Yukawa  beta function  $\beta_y$ up to two loops is given by 
\beq\label{betay}
\begin{array}{rcl}
\beta^{(1)}_y&=&
\displaystyle
  (13 + 2 \eps) \,\alpha^2_y-6\,\alpha_y\,\alpha_g \,, \\[1ex]
\beta^{(2)}_y&=&
\0{20 \eps-93}{6}\alpha_g^2 \,\alpha_y 
+  (49 + 8 \eps) \alpha_g \alpha_y^2
\\[2ex] && \displaystyle
-  4\big[(11 + 2 \eps) \alpha_y- \alpha_u\big]\,\alpha_u \alpha_y 
\\[1ex] && 
-    \left(\0{385}{8} + \0{23}{2} \eps + \0{\eps^2}{2}\right) \alpha_y^3 
\,.
  \end{array}
\eeq
The Yukawa beta function  depends on the gauge and Yukawa couplings, at any loop order. From two loop level onwards, it also depends on the scalar coupling $\alpha_u$. In the Veneziano limit, neither \eq{betag} nor \eq{betay} depends on the double-trace scalar coupling $\alpha_v$, at any loop order. 

The  beta function for the single trace scalar quartic coupling $\beta_u$ up to two loops  is given by
  \beq\label{betau}
\begin{array}{rcl}
 \beta^{(1)}_u&=&
 -(11+ 2\eps) \,\alpha_y^2+4\alpha_u(\alpha_y+2\alpha_u)\,,\\[1ex]
 \beta^{(2)}_u&=&
\alpha_u \alpha_y \big[ 10 \alpha_g - 16 \alpha_u - 3(11 + 2 \epsilon)\alpha_y\big]
\\[2ex] && 
   +(11+2 \epsilon)\big[
   (11+2 \epsilon)\alpha_y
 -2 \alpha _g 
 \big] \alpha _y^2
\\[2ex] && 
 -24 \alpha _u^3\,.
 \end{array}
\eeq  
The beta function $\beta_v$ for the double trace quartic scalar coupling is given by
\beq\label{betav}
\begin{array}{rcl}
  \beta^{(1)}_v&=&
12 \alpha_u^2  +4\al v \left(\alpha_v+ 4 \alpha_u+\alpha_y\right)\,,\\[1ex]
 \beta^{(2)}_v&=&
 8\alpha _v \alpha _y
 \big[ 
 \054 \alpha _g 
 -4 \alpha _u 
   -\alpha _v
  -(\0{33}{8}+\034 \epsilon)  \alpha _y
 \big]
\\[2ex] && \displaystyle
 +(11+2\eps)\big[(11+2\eps)  \alpha _y  + 4\alpha _u \big] \alpha _y^2
\\[2ex] && \displaystyle
    - 8\alpha _u^2
  \big[
  12\alpha_u
 +5\alpha _v 
 +3 \alpha  _y
 \big]\,.\ \ \ \ \ \ 
\end{array}
\eeq
Starting from the two loop level, both scalar beta function additionally depend on the gauge coupling. Our result is also in accord with the findings of \cite{Pomoni:2008de} which state that $\beta_v$ is quadratic in $\alpha_v$ to all loop orders in the Veneziano limit. 

Some of the expressions have previously been given in \cite{Litim:2014uca}. The main new additions here are the 2-loop scalar terms in \eq{betau} and \eq{betav}. In the Veneziano limit, the subsystem $(\beta_g,\beta_y)$ is independent of $(\alpha_u,\alpha_v)$ at the leading non-trivial order which is two (one) loop in the gauge (Yukawa, scalar) couplings. Beyond this order,  the subsystem $(\beta_g,\beta_y,\beta_u)$  remains independent of $\alpha_v$.

\begin{table*}
\begin{tabular}{cc c cc cc cc c}
 \toprule
 \rowcolor{Yellow}
 {}\ \ \ \bf Couplings\ \ \ 
 &\multicolumn{8}{c}{\ \ \bf Orders in perturbation theory\ \ }
 &{\bf  \ \ Scheme\ \ }
 \\
\midrule
\rowcolor{LightGray}     
 $\bm{\beta_{\rm gauge}}$ &1&1&2&2&2&3&3& 3&\cellcolor{white}\\
\rowcolor{white}
 $\bm{\beta_{\rm  Yukawas}}$&0&1&1&1&2&2&2&3&\\
\rowcolor{LightGray}
 $\bm{\beta_{\rm   quartics}}$ &0&1&0&1&2&1&2&3&\cellcolor{white}\\
\midrule
\rowcolor{LightGray}     
\cellcolor{white}
&& \ LO\ &&&NLO&&&2NLO&\bf PT\\
\cellcolor{white} &LO${}'$& &&\ NLO${}'$\ &&&2NLO${}'$&&\bf FP\\
\rowcolor{LightGray}     
\cellcolor{white} &\ LO${}''\ $& &\ NLO${}''$\ &&&2NLO${}''$&&&\bf Weyl\\
\bottomrule
\end{tabular}
\caption{\label{Tab} 
Approximation schemes sorted according to the loop orders retained in the various beta functions, comparing perturbation theory (PT),   fixed point consistency conditions  (FP) \cite{Litim:2014uca,Litim:2015iea}, and Weyl consistency conditions (Weyl), each to leading (LO), next-to-leading (NLO) and next-to-next-to-leading (2NLO) order.}\label{tNLO}
\end{table*} 

\subsection{Anomalous dimensions}\label{anomalous}
We also provide results for the anomalous dimensions associated to the fermions and scalars \cite{Machacek:1983tz,Luo:2002ti}. If mass terms are present, their renormalisation group flow is induced through the RG flow of the gauge, Yukawa, and scalar couplings. Following  \cite{Bond:2017lnq}, we define the scalar anomalous dimensions as
$\Delta_H=1+\gamma_H$, where $\gamma_H\equiv  \frac{1}{2}{d\ln Z_H}/{d\ln\mu}$, and the fermion anomalous dimension as $\gamma_Q\equiv {d\ln Z_Q}/{d\ln\mu}$. 
Within perturbation theory, the one and two loop contributions read
\beq
\label{gamma}
\begin{array}{rcl}
\gamma_H&=&
\alpha_y
-\032\left(\0{11}{2}
-\eps\right)\alpha_y^2
+\052\alpha_y\,\alpha_g
+2\alpha_u^2 \,,\\[2ex]
\gamma_Q&=&
\left(\0{11}{2}+\eps\right)\alpha_y
+\xi\,\alpha_g \\[1ex]
&&- \left(\eps - 2\xi -\014 \xi^2 \right)\,\alpha_g^2 -\left(11 + 2 \eps\right)\,\alpha_g\,\alpha_y
\\[1ex]
&&
-\left(\0{253}{16} + \0{17}{4}\eps + \014 \eps^2 \right)\,\alpha_y^2\,,
\end{array}
\eeq
up to corrections of order ${\cal O}(\alpha^3)$. Here, $\xi$ denotes the $R_\xi$ gauge fixing parameter. The anomalous dimension for the scalar mass term follows from the composite operator $\sim M^2\,\Tr\, H^\dagger H$ with
$\gamma_{M}=  d\ln M^2/d \ln \mu$.  
The anomalous dimension for the fermion mass operator is defined as
$\Delta_Q=3+\gamma_{M_Q}$ with $\gamma_{M_Q}\equiv  {d\ln M_Q}/{d\ln\mu}$. Within perturbation theory, we find
\beq\label{gammaM}
\begin{array}{rcl}
\gamma_{M}&=&8\al u+4 \al v+2\al y - \left(\frac{33}{2} + 3\eps\right)\,\alpha_y^2 
\\[1ex] &&
- \left( 16 \alpha_u + 8 \alpha_v -5 \alpha_g \right)\alpha_y 
- 20 \,\alpha_u^2\,, \\[2ex]
\gamma_{M_Q}&=&
\left(\0{11}{2}+\eps\right)\alpha_y-3\, \alpha_g  
+ \left(22 + 4\eps\right) \alpha_g\,\alpha_y 
\\[1ex] &&
- \left(\0{31}{4} - \053 \eps\right)\alpha_g^2 
-\left(\0{253}{16} + \0{17}{4}\eps + \0{\eps^2}{4}\right) \alpha_y^2
\end{array}
\eeq
up to terms of order ${\cal O}(\alpha^3)$. We note that $\gamma_{M}$ is manifestly positive at leading order. For $\gamma_{M_Q}$ we observe that the gauge and Yukawa contributions arise with manifestly opposite signs at leading order. Hence these  may take either sign respectively, depending on whether the gauge or Yukawa contributions dominate.  Utilizing \eq{gammaM}, mass terms then evolve according to
\beq
\label{gammaF}
\begin{array}{rcl}
\beta_{M^2} &=& \gamma_{M}\, M^2-8\,\alpha_y\, M_Q^2
+ {\cal O}(\alpha^3)\,, \,\\[1ex]
	\beta_{M_Q} &=& \gamma_{M_Q}\,M_Q+ {\cal O}(\alpha^3) \,.
\end{array}
\eeq
The flow of mass terms already mixes to leading order in the couplings, even in the Veneziano limit. Additional mixing contributions are present as soon as $N_C$ and $N_F$ take finite values.

\subsection{Systematics}
Next, we discuss the systematics of fixed point searches in perturbation theory, Tab.~\ref{Tab}. Our considerations in this section apply to any $4d$ theory with weakly-coupled fixed points, and are more general as such than the concrete asymptotically safe model introduced above.

Theories in $4d$ without gauge interactions cannot develop weakly coupled fixed points \cite{Bond:2016dvk,Bond:2017sem}. Hence, gauge interactions must invariably be present to generate fixed points at weak coupling. Scalar or Yukawa couplings may also be present, depending on the particulars of the matter content. If so, scalar quartic and Yukawa couplings and their beta functions arise alongside those for the gauge couplings. We then denote  the approximations which retain
terms up to order $k$, $n$, and $m$ in the loop expansion of the gauge, Yukawa, and scalar beta functions by
\beq
\label{kmn}
(k,m,n)\,.
\eeq
Whenever unambiguous, we drop the commas inbetween.
  Evidently, without scalars, we have $n=m=0$ throughout. One might wonder which approximation orders lead to self-consistent fixed points.

Within perturbation theory, and without any other a priori information about the theory,
it seems natural to retain beta functions up to the same loop order for all couplings, corresponding to the sequence
\beq\label{PT}
{\rm PT:}\quad (n,n,n)\,.
\eeq
The first few approximations are the leading order (111), the next-to-leading order (222), and the next-to-next-to-leading order (333), as indicated in Tab.~\ref{Tab}. 

In theories with weakly interacting fixed points, however, further information is available. In fact, close to fixed points the naive perturbative ordering is upset owing to interactions. It has been established in  \cite{Bond:2016dvk,Bond:2017sem} that any weakly interacting fixed point requires the one loop gauge coefficient to be parametrically small.\footnote{Strictly speaking, it is required that the ratio of the one-loop and the two-loop gauge coefficient is a perturbatively small number. If so, it can then always be achieved that the gauge one loop coefficient is small by a suitable reparametrisation of the gauge coupling.}    If we denote the  small parameter which controls the smallness of the gauge one loop coefficient by $\eps$ (in the model \eq{betag}, the one-loop coefficient  reads $-\s043\eps$),  this structure implies that
$\beta^{(1)}_g\sim \eps\, \alpha^2_g\ll  \alpha^2_g$.
In such settings, the leading order approximation is (100) rather than (111) owing to the parametric slowing-down of the gauge coupling as opposed to the other sectors. 
  Barring exceptional cancellations, this structure also implies that the one and two loop gauge contributions are of the same order of magnitude
  $
  \beta^{(1)}_g\sim \beta^{(2)}_g \sim \eps^3$,
 close to interacting fixed points $\alpha^*\sim \eps$, see \eq{betaPT}.  On the other hand,  Yukawa and scalar beta functions at one loop cannot be made parametrically small. Consequently, the   approximation which provides the first  order at which  a consistent fixed point $\alpha^*={\cal O}(\eps)$  for all couplings arises is (211): in the gauge sector the fixed point materialises due to cancellations between the one and two loop terms, and in the Yukawa and scalar sectors through cancellations at one loop  \cite{Bond:2016dvk,Bond:2017sem}.  
  All higher loop contributions are parametrically smaller and obey $\beta^{(n)}\sim \eps^{n+1}$ for the gauge beta function once $n\ge 2$ as well as $\beta^{(n)}\sim \eps^{n+1}$ for the Yukawa and the scalar beta functions for all $n\ge 1$. This pattern proceeds systematically to higher order \cite{Bond:2017lnq}. 
It follows that the sequence of approximations with consistent interacting fixed point (FP) solutions is given by 
\beq\label{FP}
{\rm FP:}\quad (n+1,n,n)\,.
\eeq
We denote this approximation as $n$NLO${}^\prime$. 
 It determines the fixed point $ \alpha^*(\eps)=\alpha^*|_{n{\rm NLO'}}
+{\cal O}(\eps^{n+1})$
for all couplings, with $\alpha^*|_{n{\rm NLO'}}$ an exact polynomial in $\eps$ up to including terms of order $\eps^n$.
The first few approximations are the leading (100), the next-to-leading (211), and the next-to-next-to-leading (322) order, see Tab.~\ref{Tab}. 

Finally, a third sequence of approximations exploits information  related to Weyl consistency conditions \cite{Jack:1990eb,Jack:2013sha}. Weyl consistency conditions have formally been derived for weakly coupled theories on classical gravitational backgrounds. On the level of the path integral they state that two independent Weyl rescalings commute with each other.  
In terms of the couplings  $\{ g_i \} \equiv \{ g, y,u,v\}$ with $\beta$ functions $\beta_i=dg_i/d\ln\mu$,
the Weyl consistency conditions take the form of integrability conditions
	${\partial \beta^j}/{\partial g_i} = {\partial \beta^i}/{\partial g_j}$ in that they
relate partial derivatives of the various $\beta$ functions to each other, and $\beta^i \equiv \chi^{ij} \beta_j$. The functions $\chi^{ij}$ play the role of a metric in the space of couplings. Weyl consistency conditions are expected to hold in the full theory, and hence it might seem desirable to satisfy them even within finite perturbative approximations. Note that the metric $\chi^{ij}$ itself is a function of the couplings which is why Weyl-consistent solutions  relate different orders of perturbation theory. For the gauge-Yukawa theory studied here, a perturbative expression for the metric $\chi$ has been given in \cite{Antipin:2013pya}. Accordingly, Weyl-consistent approximations are given by the sequence
\beq\label{Weyl}
{\rm Weyl:}\quad (n+1,n,n-1)\,.
\eeq
We denote this approximation as $n$NLO${}''$.  The first few approximations are the leading (100), the next-to-leading (210), and the next-to-next-to-leading (321) order, see Tab.~\ref{Tab}. Notice that the FP \eq{FP} and Weyl \eq{Weyl}  approximations only differ in the scalar sector, where the former retains an additional loop order. However, in any QFT, scalar couplings only enter the Yukawa beta functions starting at two loop order, and the gauge sector at even higher loop level. For this reason, the higher loop term in the scalar sector only generates subleading corrections for the gauge and Yukawa fixed point. This pattern implies that power series expansions of fixed points at  $n$NLO${}^\prime$ or  $n$NLO${}''$ accuracy
coincide for the gauge, Yukawa (scalar) couplings, modulo subleading terms of order $\sim\eps^{n+1}$  ($\sim\eps^{n}$), for all $n$.

The PT and Weyl schemes up to 2NLO and 2NLO${}''$ have recently been used to investigate the vacuum stability of the Standard Model \cite{Degrassi:2012ry,Antipin:2013sga}.  For the model at hand \eq{L}, \eq{Lall}, the approximations 
NLO${}''$, NLO${}'$, and 2NLO${}''$
have been investigated in \cite{Litim:2014uca,Litim:2015iea}. Below, we extend approximations  to the complete 2NLO${}^\prime$ order (322) in  the spirit of \eq{FP}, and compare the PT, FP, and Weyl approximation schemes quantitatively.

\subsection{Away from four dimensions}

As an aside, we note that the power counting detailed in Tab.~\ref{Tab} applies uniquely to weakly interacting QFTs in $4d$. Away from four dimensions, the gauge, Yukawa and quartic self-interactions have a non-vanishing canonical mass dimension, and their $\beta$-functions receive a tree level contribution which alters the power counting in Tab.~\ref{Tab}. Specifically, in $d=4-\delta$ dimensions, the tree level parameter $|\delta|\ll 1$ now controls the perturbative expansion and the existence of fixed points.   
Barring exceptional cancellations, 
the leading non-trivial order with a consistent interacting fixed point $\alpha_i^*={\cal O}(\delta)$ is one-loop $(111)$, where quantum fluctuations cancel the tree level terms for some or all couplings (see \cite{Codello:2016muj} for a recent example).
This pattern proceeds to higher order, as is well-known from, e.g., the  Wilson-Fisher fixed point \cite{Wilson:1971dc}.

\section{\bf Results at 2NLO${}^\prime$}\label{2NLO}
In this section, we summarise our results for fixed points, anomalous dimensions, vacuum stability, and scaling exponents at the complete 2NLO${}^\prime$ order. 
\subsection{Fixed points}

It is straightforward if tedious to identify the weakly interacting fixed points at order $\eps^2$ of the system \eq{betag}, \eq{betay}, \eq{betau} and \eq{betav}. Given the polynomial nature of the beta function, however, a large variety of (potentially spurious) fixed points arises. Those fixed points which are proportional to $\eps$ in the leading order are under strict perturbative control and can be viewed as ``exact''.  Using the beta functions at (322) accuracy, and performing a systematic expansion \eq{FP} up to subleading corrections of order $\eps^3$, we find 
\beq\label{alpha2NLO}
\begin{array}{rcl}
\alpha_g^*&=&
\displaystyle
\0{26}{57}\,\eps
+ 23\0{75245 - 13068 \sqrt{23}}{370386}\,\eps^2\,,
\\[2ex]
\alpha_y^*&=&
\displaystyle
\0{4}{19}\,\eps
+\0{43549-6900 \sqrt{23}}{20577}\,\eps^2 \,,
\\[2ex]
\alpha_u^*&=&
\displaystyle
\0{\sqrt{23}-1}{19}\,\eps
+\0{365825\sqrt{23}-1476577}{631028}\,\eps^2\,, 
\\[2ex]
\al {v}^*&=&
\displaystyle
-\0{1}{19} \left(2 \sqrt{23}-\sqrt{20 + 6 \sqrt{23}}\right)\,\eps
 \\[2ex] && \displaystyle
-\left(
\0{321665}{13718\sqrt{23}}
-\0{27248}{6859}
+\0{\0{33533}{6859}-\0{452563}{13718\sqrt{23}}}{\sqrt{20+6\sqrt{23}}}
\right)\eps^2\,.
\end{array}
\eeq
Results are accurate at the cited order, meaning that higher loop corrections will only generate subleading terms of order $\eps^3$. Results agree with the (321) approximation adopted previously \cite{Litim:2014uca} in all but the $\eps^2$-terms of the scalar quartic couplings.  The reason for this is that the scalar couplings interfere with the Yukawa and gauge beta functions starting at the second and fourth loop level, respectively, see \eq{betay}. In consequence, at (322), only the $O(\eps)$ coefficent of the scalar couplings contribute to the $O(\eps^2)$ value of the Yukawa coupling, whence agreement with (321). Quantitatively, we have
\beq\label{alpha2NLOnum}
\begin{array}{rcl}
\alpha_g^* &= &\ \ \,0.4561 \epsilon 
+0.7808 \epsilon ^2 
+3.8922 \epsilon ^3
\,,\\[1ex]
\alpha_y^* &= &\ \ \,0.2105 \epsilon +0.5082 \epsilon ^2 
+2.4222 \epsilon ^3
\,, \\[1ex]
\alpha_u^* &=&\ \ \,0.1998 \epsilon +0.4403 \epsilon ^2
+1.8780 \epsilon ^3
\,, \\[1ex]
\alpha_v^* &=&-0.1373 \epsilon -0.6318 \epsilon ^2
-3.6685 \epsilon ^3\,.
\end{array}
\eeq
All terms have coefficients of order unity. We have also indicated the $\eps^3$ terms which originate from subleading contributions in $\eps$ at 2NLO${}^\prime$ accuracy; they are only indicative as further higher loop corrections beyond (322) will modify them.  Also note that all terms at order  $\eps^2$ arise with the same sign as those at order $\eps$. This implies that the $\alpha_g$  and $\alpha_y$ remain positive for all $\eps$, as they must, offering no limitations on the domain of validity.  
It would be very useful to know whether the radius of convergence (in $\eps$) comes out finite, or not.  Same sign correction terms hint at a slow rate of convergence in $\eps$ and the presence of complex conjugate poles in the complexified  field plane \cite{Litim:2016hlb,Juttner:2017cpr}.

\subsection{Vacuum stability}

We now turn our attention to the stability of the vacuum. It is well-known that  the scalar couplings  control the stability of the ground state. The stability for scalar potentials as in \eq{L} has first been investigated in \cite{Paterson:1980fc}.  In the Veneziano limit, and in terms of the couplings used here, it is required that \cite{Litim:2014uca,Litim:2015iea}
\beq
\label{stability}
\alpha_u^*>0\quad {\rm and}\quad\alpha_u^*+\alpha_v^*>0\,.
\eeq
The first approximation with non-trivial scalar couplings is NLO${}^\prime$ (211).  At one loop, the fixed point in the scalar sector is fuelled by the Yukawa fixed point. Most importantly, vacuum stability has been established quantitatively \cite{Litim:2014uca}, with
\beq
\label{stability211}
\alpha_u^*+\alpha_v^*\,\Big|_{(211)}=0.0625 \eps +O\left(\epsilon ^2\right)\,.
\eeq
Notice the smallness of the leading coefficient. It arises through the cancellation of the leading order fixed point values of the single and double trace couplings, which by themselves are twice or thrice as large as their sum, \eq{alpha2NLO}. It has also been shown that the Coleman-Weinberg-type resummation of leading logarithmic corrections does not alter the conclusion \cite{Litim:2015iea}. A first step beyond the leading order (211) has been performed in \cite{Litim:2014uca} by using the Weyl-consistent (321) approximation. The result 
\beq
\label{stability321}
\alpha_u^*+\alpha_v^*\,\Big|_{(321)}=0.0625 \eps + 0.1535 \eps^2
\eeq
shows that the induced subleading, higher loop effects from the gauge-Yukawa sector are supportive of vacuum stability, for all $\eps$.  This can also be understood from observing that the scalar quartic couplings are at one loop proportional to the Yukawa coupling; and since the latter grows with subleading corrections, so does \eq{stability321} over \eq{stability211}. At (322) accuracy, however, we find the complete $\eps^2$-correction from \eq{alpha2NLO}. Quantitatively, we have
\beq
\label{stability322}
\alpha_u^*+\alpha_v^*\,\Big|_{(322)}=0.0625 \eps - 0.1915 \eps^2+O\left(\epsilon ^3\right)\,.
\eeq
Notice that the leading and the subleading terms now arise with opposite signs. At order $\eps^2$, this comes about because the double-trace scalar coupling receives larger (and negative) corrections than the single trace coupling.  We also observe that the two loop terms in the scalar beta function  outweigh the gauge-Yukawa corrections in \eq{stability321}.

\subsection{Anomalous dimensions}
For the field and mass anomalous dimensions, using \eq{gamma} and \eq{gammaM} in conjunction with \eq{alpha2NLO}, we find 
\beq
\begin{array}{rcl}
\gamma_H\big|_{(322)}&=&\ \ 0.211\,\eps + 0.462\,\eps^2\,,\\[1ex]
\gamma_Q\big|_{(322)}&=&\ \ (1.158 + 0.456\,\xi)\,\eps \\
&&+ (1.249 + 1.197\,\xi + 0.052\,\xi^2)\,\eps^2\,,\\[1ex]
\gamma_{M}\big|_{(322)}&=&\ \ 1.470\,\eps + 0.521\,\eps^2\,,\\[1ex]
\gamma_{M_Q}\big|_{(322)}&=&-0.421\,\eps + 0.926\,\eps^2\,,
\end{array}
\eeq
up to terms of order ${\cal O}(\eps^3)$, and  where $\xi$ denotes the gauge fixing parameter in $R_\xi$ gauge. We observe that the subleading corrections have the same sign as the leading order ones, except for the mass anomalous dimension $\gamma_{M_Q}$.  Results are  compatible with unitarity bounds. 
With increasing $\eps$, the anomalous dimension $\gamma_{M}$ exceeds the classical dimension starting at about $\eps=1.36$ at (211) or $\eps=1.00$ at (322). For the fermion mass anomalous dimension, this happens at $\eps=1.29$ at (322). This implies that mass terms become  irrelevant operators in the UV for sufficiently large $\eps$. We interpret this phenomenon as the onset of strong coupling where the validity of perturbation theory becomes questionable.

\subsection{Scaling exponents}
Next we discuss universal exponents which are obtained as eigenvalues of the stability matrix $\partial\beta_i/\partial \alpha_j|_*$. We order the eigenvalues according to magnitude, $\vartheta_1<0<\vartheta_2<\vartheta_3<\vartheta_4$.\footnote{This relates to the convention used in \cite{Litim:2014uca} under the exchange $\vartheta_3 \leftrightarrow \vartheta_4$.}
Scaling exponents have been known at (211) and (321)  accuracy previously \cite{Litim:2014uca}.  Our results for the scaling exponents at 2NLO${}^{\prime}$ are
\beq\label{theta2NLO}
\begin{array}{rcl}
\vartheta_1&=&
\displaystyle
-\0{104}{171}\,\eps^2
+\0{2296}{3249}\,\eps^3\,,
\\[3ex]
\vartheta_2&=&
\displaystyle
\ \ \,\0{52}{19}\, \eps  + \0{136601719-22783308 \sqrt{23}}{4094823}\,\eps^2\,,
\\[3ex]
\vartheta_3&=&
\displaystyle
 \ \ \0{8}{19} \sqrt{20 + 6 \sqrt{23}}\,\eps 
 \\[2ex] && \displaystyle
 + \0{2 \sqrt{2} \left(50059110978+10720198219 \sqrt{23}\right)}{157757 \left(10+3 \sqrt{23}\right)^{9/2}}\,\eps^2\,,
\\[3ex]
\vartheta_4&=&
\displaystyle
\ \ \0{16}{19}\sqrt{23}\,\eps
+ \0{4 \left(68248487 \sqrt{23}-255832864\right)}{31393643}\,\eps^2
\,.
\end{array}
\eeq
The new coefficients are $\eps^2$-corrections to the irrelevant eigenvalues $\vartheta_3$ and $\vartheta_4$, which are the only terms sensitive to the two-loop scalar beta functions; the $\sim \eps^2$ contribution to $\vartheta_2$ is only dependent on the scalar couplings to one-loop.
Note that the rational coefficients in $\vartheta_1$ and $\vartheta_2$ arise from the gauge-Yukawa subsector, whereas all irrational coefficients arise with contributions from the scalar subsector. 
It is interesting to note that the relevant scaling exponent $\vartheta_1$ is completely determined to $O(\eps^3)$ already at (210) order, as noted in \cite{Litim:2014uca}. However, in contrast to the other exponents, and expectation, increasing our approximation to (322) does not fix any further coefficients, as the $\sim \eps^4$ coefficient is sensitive to four-loop (three-loop) contributions to the gauge (Yukawa) beta functions.
Numerically, we have
\beq\label{theta2NLOnum}
\begin{array}{rl}
\vartheta_1 = &-0.6082 \epsilon ^2
+0.7067 \epsilon ^3+3.322 \eps^4\,, 
 \\[.5ex]
\vartheta_2 = &\ \  2.737 \epsilon +6.676 \epsilon ^2
+18.44 \epsilon ^3\,,
\\[.5ex]
\vartheta_3 = &\ \  2.941 \epsilon +1.041 \epsilon ^2
-2.986 \epsilon ^3\,,
\\[.5ex]
\vartheta_4 = &\ \  4.039 \epsilon +9.107\epsilon ^2
+44.43 \epsilon ^3\,,
\end{array}
\eeq
where we additionally show the next subleading coefficient in each case (e.g. the $\eps^4$-term in $\vartheta_1$ and the $\eps^3$-terms for the other exponents). The latter terms are subject to corrections from the next loop level, and quantify subleading effects already present within the  (322) approximation. 

\section{\bf UV conformal window}\label{UV}
We are now in a position to investigate the size of the UV conformal window for asymptotic safety for theories with action \eq{L} using perturbation theory.

\subsection{Limits for interacting fixed points}
The results of the previous sections have established a UV fixed point to second order in $\eps\ll 1$. With increasing $\eps$, the conformal window for the UV fixed point is limited through one of several mechanisms:

\begin{itemize}
\item[${a)}$] {Strong coupling}. With increasing $\eps$, regimes with parametrically strong coupling in $\eps$ can arise either through algebraic poles of fixed point couplings $\alpha(\eps)$ at finite $\eps$, or in the limit $\eps\to \infty$. In the latter case, we impose $\alpha^*<1$ to delimit the range of validity. 

\item[${b)}$] {Fixed point mergers.}  Fixed point conditions for approximations beyond (211) are at least  quadratic (or higher) order in one of the couplings. Consequently, additional strongly coupled IR fixed point solutions may arise. With increasing $\eps$, these may collide with the asymptotically safe UV fixed point, and then  disappear in the complex plane, setting an upper limit on $\eps$. Equivalently, this is signalled by the vanishing of the relevant scaling exponent.

\item[${c)}$] {Vacuum instability.} The signs and size of the scalar couplings are solely constrained by the requirement of vacuum stability  \eq{stability}. Consequently, the change of sign for the linear combination  \eq{stability} with increasing $\eps$ indicates  the onset of instabilities. 
\item[${d)}$] {Negative coupling.}  Regions with parametrically weak gauge or Yukawa coupling $\alpha(\eps)\to 0$  for increasing $\eps>0$ offer upper limits due to a change of sign of these couplings and the subsequent disappearance of fixed points into the unphysical regime.
\end{itemize}
From the point of view of practical  applications,  it is crucial to understand up to which finite maximal value $\eps< \eps_{\rm max}$ the conformal window is going to persist, and which mechanism is responsible for generating an upper bound, if any.

\begin{figure}[t]
\begin{center}
\hskip-.5cm
\includegraphics[scale=.9]{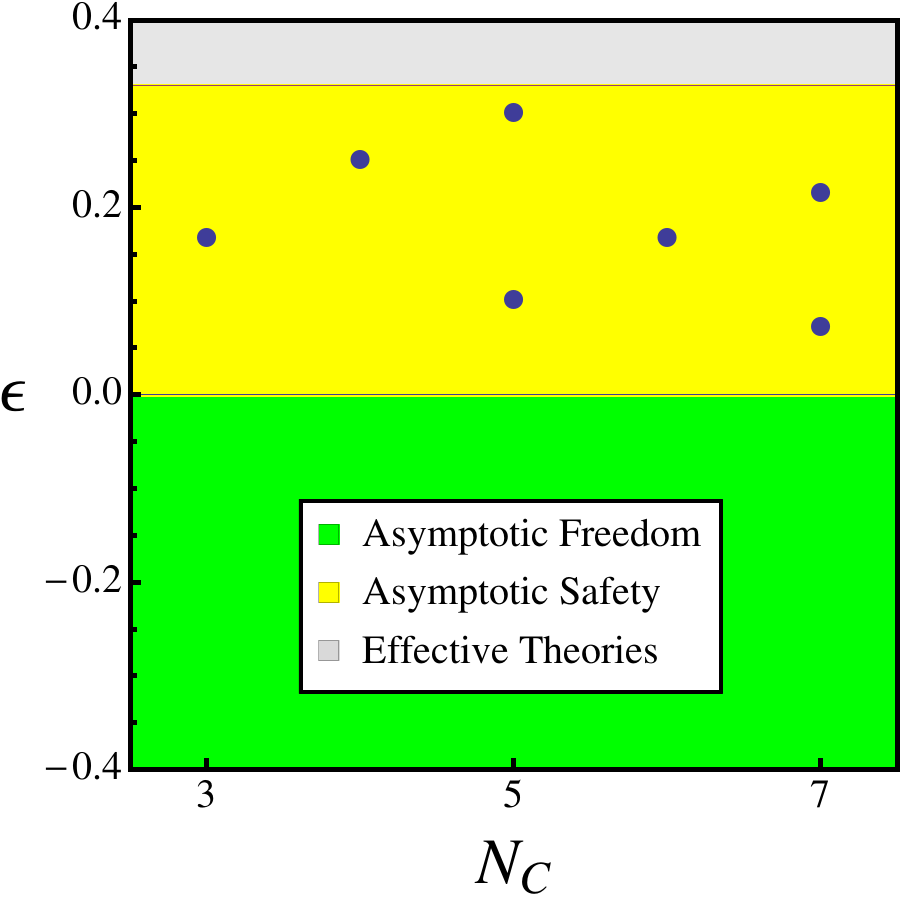}
\caption{\label{pWindow1}
The UV conformal window with asymptotic safety (yellow band) from fixed points and scaling exponents, \eq{max}, also showing regimes with asymptotic freedom (green) and effective theories (grey). Dots indicate the first few integer solutions \eq{maxN}.}
 \vskip-.6cm
 \end{center}
\end{figure}

\subsection{Bounds from fixed points and exponents}

A first estimate for an upper bound  follows from the complete results at (211) and (322) order for the couplings (up to second order in $\eps$), and the scaling exponents (up to fourth order in $\eps$).  Since all  couplings receive same-sign corrections at (322), \eq{alpha2NLO},  the scenario $d)$ cannot arise. Requiring $\alpha^*<1$ leads to $\eps< 2$ approximately. However, vacuum stability offers tighter constraints. We conclude from  \eq{stability322} that the two loop scalar corrections  impose an upper bound for the conformal window  through the onset of vacuum instability, approximately given by $\eps_{\rm max}\approx 0.326$.

Let us see whether some of the incomplete higher order corrections offer a similar, or even tighter bound.
From the relevant eigenvalue \eq{theta2NLO}, an upper limit $\eps_{\rm max}\approx 0.861$ arises from sign change of $\vartheta_1$ through the incomplete $\eps^3$ term, indicating a fixed point merger \cite{Litim:2014uca}. 
Considering incomplete $\eps^4$ contributions from (322), the upper bound is reduced to $\eps_{\rm max}\approx 0.335$. A sign change in $\vartheta_3$ would arise at even larger $\eps$ and can be ignored.  No constraints arise from anomalous dimensions. 
Based on the explicit power series expressions for couplings and exponents at 2NLO${}'$, we conclude that the conformal window is limited  through the onset of vacuum instability \eq{stability322} and the vanishing of the relevant eigenvalue \eq{theta2NLO},
\beq\label{max}
\eps_{\rm max}\approx 0.326\, \dots \,0.335\,,
\eeq
see Fig.~\ref{pWindow1}. 
It is interesting to observe that the tightest bound from incomplete higher order terms comes out very close to (yet, larger than) the vacuum stability bound. In this light, we view \eq{max} as indicative for the range of validity at this order. 
Constraints through parametrically strong or weak coupling do not play any role.  As we will see next, the UV conformal window becomes more strongly constrained once bounds from beta functions are taken into consideration.

\begin{table*}[t]
\begin{tabular}{cc cc cc cc cc}
 \toprule
 \rowcolor{Yellow}
 {}\ \ \ \bf Couplings\ \ \ 
 &\multicolumn{9}{c}{\ \ \bf Orders in perturbation theory\ \ }
 \\
\midrule
\rowcolor{LightGray}     
 $\beta_{\rm gauge}$ &2&2&2&2&2&3&3&3& 3\\
\rowcolor{white}
 $\beta_{\rm  Yukawas}$&1&1&1&2&2&1&1&2&2\\
\rowcolor{LightGray}
 $\beta_{\rm   quartics}$ &0&1&2&1&2&1&2&1&2\\
\midrule
$\bm{\eps_{\rm strict}}$
&$2.192^a$
&$2.192^a$
&$0.135^c$
&$16.16^a$
&$0.222^c$
&$0.029^b$
&$0.029^b$
&$0.145^b$
&$0.095^c$\\
\rowcolor{LightGray}
$\bm{\eps_{\rm subl.}}$
&$\ 1.048^a\ $
&$\ 1.048^a\ $
&$\ 0.116^c\ $
&$\ 3.112^b\ $
&$\ 0.208^c\ $
&$\ 0.027^b\ $
&$\ 0.027^b\ $
&$\ 0.117^b\ $
&$\ 0.087^c\ $\\
\bottomrule
\end{tabular}
\caption{\label{TabEps} Maximal values
$\eps_{\rm strict}$ and $\eps_{\rm subl.}$
for the parameter $\eps$ up until which asymptotic safety is realised.
Limits arise due to $a)$ strong coupling, $b)$ fixed point mergers, or $c)$ vacuum instability.}\label{tmax}
\end{table*}

\subsection{Stabilising vs destabilising fluctuations}

Next, we investigate constraints arising directly from the beta functions rather than  their power series solutions. We will see that this leads to tighter constraints yet.  
As a first step, 
it is interesting to ask into which direction the higher loop corrections are going to shift the beta functions. Inserting the order $\eps$ fixed point results from \cite{Litim:2014uca} 
into the higher loop terms, 
we find  the leading shifts
\beq\label{shift}
\begin{array}{l}
\beta^{(3)}_g\big|_{(211)}=2.48\,\eps^4\,,\quad
\beta^{(2)}_y\big|_{(211)}=-0.49\,\eps^3\,,\\[2ex]
\beta^{(2)}_u\big|_{(211)}=0.26\,\eps^3\,,\quad
\beta^{(2)}_v\big|_{(211)}=0.99\,\eps^3\,.
\end{array}
\eeq
Higher loop contributions to the gauge (Yukawa, scalar) sectors do not appear until order $\eps^4$ ($\eps^3)$, as is necessarily the case.
At the leading non-trivial order in $\eps$, the fixed point at the leading order (211)  shifts  the subleading gauge and scalar beta functions upwards, but the Yukawa beta function  downwards, see \eq{shift}. In general, upward shifts $\Delta \beta>0$ at some finite couplings  potentially destabilise UV fixed points, simply because beta functions might no longer be able to generate a non-trivial zero once upward shifts become too large. For the same reason, downward shifts $\Delta \beta<0$ always stabilise interacting UV fixed points, simply because $\beta>0$ for sufficiently small couplings, which guarantees that a solution to $\beta = 0$ can still be found for finite positive couplings. Altogether this means that  higher loop corrections \eq{shift} to the running of the Yukawa (gauge, scalar) coupling  stabilise (de-stabilise) the fixed point. It remains to be seen how this ``competition of fluctuations'' balances out quantitatively across the various beta functions and loop orders.

\subsection{Bounds from beta functions}
Next, we determine bounds from beta functions quantitatively \cite{Medina:2016}. We  adopt two strategies to determine $\eps_{\rm max}$ from beta functions, for each set of loop orders.
The first ``strict'' strategy, whose bounds we call 
$\eps<\eps_{\rm strict}$, 
uses the loop orders as indicated in Tab.~\ref{TabEps}. In  addition, 
all terms in the beta functions \eq{betaPT} which are parametrically larger than  $\eps^{n+1}$ at the $n$-th loop order are suppressed (couplings count as $\alpha\sim \eps$).  
The rationale for this strict approach is that the approximate beta functions  are now stripped of those higher order contributions (in $\eps$), which are not (yet) accurately determined due to  the absence of higher loop terms. As such, the scheme primarily acknowledges the power counting $\alpha\sim \eps$, as dictated by the fixed point. The bounds $\eps_{\rm strict}$ are  sensitive to the competition between the stabilising Yukawa and the destabilising gauge and scalar loop contributions at higher order \eq{shift}.

The second strategy is agnostic to these finer considerations and employs the plain loop level approximation as discussed in Tab.~\ref{TabEps}, without touching the explicit $\eps$ dependence within loop coefficients. This strategy retains subleading terms in $\eps$ and we refer to its bounds as $\eps_{\rm subl.}$.  With the result \eq{alpha2NLO} at hand, we can estimate what the effect of these subleading terms is going to be by
inserting the fixed point solutions to order $\epsilon^2$ back into the beta functions at (322), finding
\beq\label{shift322}
\begin{array}{l}
\beta_g\big|_{(322)}=10.24\,\eps^5\,,\quad
\beta_y\big|_{(322)}=-1.71\,\eps^4\,,\\[2ex]
\beta_u\big|_{(322)}=1.70\,\eps^4\,,\ \  \quad
\beta_v\big|_{(322)}=7.24\,\eps^4\,.
\end{array}
\eeq
Subleading terms contribute starting at order $\eps^5$ ($\eps^4$) in the gauge (Yukawa, scalar) sectors, as expected from \eq{alpha2NLO}. Most notably, we find that the subleading terms shift the gauge and scalar beta functions upwards and the Yukawa beta function downwards. This is the exact same pattern as observed in \eq{shift}, albeit smaller by a power in $\eps$. Moreover, once $\eps\approx 0.14$  $(0.25)$, the scalar (gauge, Yukawa) shifts \eq{shift322} are of the same size as \eq{shift}. Since  the bounds $\eps_{\rm subl.}$ are  sensitive to the combined effect of \eq{shift} and \eq{shift322}, our line of reasoning suggests that the bounds $\eps_{\rm subl.}$ must follow the same pattern as   $\eps_{\rm strict}$ albeit being slightly tighter due to the additional shift \eq{shift322}. 

In Tab.~\ref{tmax} we summarise results for $\eps_{\rm strict}$ and $\eps_{\rm subl.}$, also indicating which mechanism is limiting the domain of validity for each case.  At the lowest orders (210), (211) and (221), we observe that $\eps_{\rm strict}$ is constrained via $\alpha^*<1$.
At (221), mergers in the Yukawa sector could have arisen.  However, the growth of the coupling with $\eps$ is much slower due to a large negative quadratic correction, $\alpha_g = 0.456 \eps -3.061 \eps^2+ O(\eps^3)$,
leading to a wider UV conformal window and the avoidance of mergers.
In (210) and (211) bounds for $\eps_{\rm subl.}$ arise from the onset of strong coupling through a pole at finite $\eps$. In these cases, the effective gauge two loop coefficient changes sign and findings can no longer be trusted in perturbation theory. At (221), instead, the bound for $\eps_{\rm subl.}$ arises through a proper fixed point merger.  As soon as two-loop effects in the scalar sector are retained, such as in (212) and (222), we find that the onset of vacuum instability dominates the upper limit. Quantitatively, the bounds are weaker in (222) than in (212). Hence, two loop Yukawa (scalar) terms increase (decrease)
the domain of validity and the conformal window.

    Turning to three loop effects, we observe that (311) is limited by fixed point mergers through fluctuations in the gauge sector. The new effect is triggered by a large positive quadratic correction $\alpha_g = 0.456 \eps +3.841 \eps^2  
    + O(\eps^3)$ which accelerates the growth of the gauge coupling, the exact opposite of what happens in (221). The effect clearly dominates over the bounds found at the preceeding orders (210), (211) and (221). 
    This continues to be true at (312), where gauge fluctuations offer a tighter constraint than vacuum stability. Including two loop Yukawa contributions, however, we find that the domain of validity is substantially enhanced --- by a factor of four in (321) and a factor of about three in (322). While in (321) the upper limit arises due to mergers,  in (322) it comes about through vacuum instability.

\begin{figure}
\begin{center}
\hskip-.5cm
\includegraphics[scale=.9]{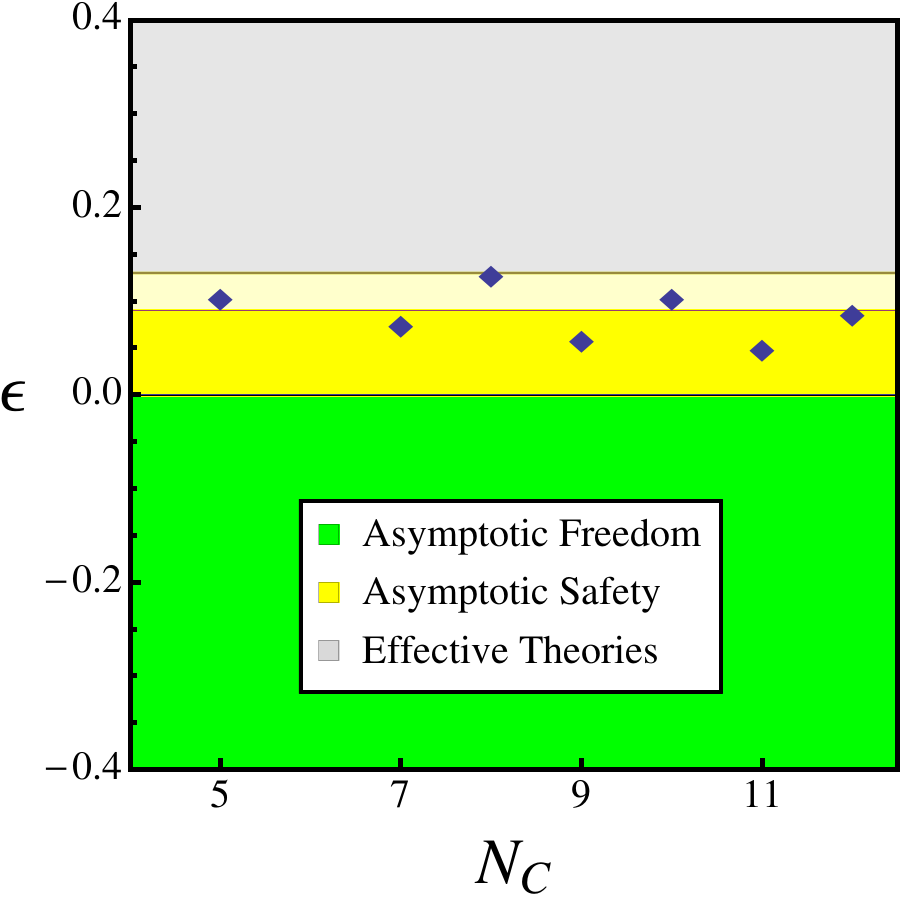}
\caption{\label{pWindow}
The UV conformal window with asymptotic safety (yellow bands) from beta functions, also showing regimes with asymptotic freedom (green) and effective theories (grey).  The lower yellow band corresponds to the full 2NLO${}'$ result, the upper yellow band covers the range \eq{epsbeta}, and symbols indicate the first few integer solutions \eq{321} and \eq{322}.}
 \vskip-.6cm
 \end{center}
\end{figure}

We now return to the induced shifts \eq{shift} and \eq{shift322}. From  Tab.~\ref{tmax}, and for all settings considered, it is evident  that the bound $\eps_{\rm subl.}$ is systematically tighter than the bound $\eps_{\rm strict}$, 
\beq\label{compare}
\eps_{\rm subl.}\klgl\eps_{\rm strict}\,.
\eeq
The result thus 
validates our semi-quantitative considerations based on induced shifts of beta functions, see \eq{shift} and \eq{shift322}. 
 We will now discuss our results from the viewpoint of perturbation theory \eq{PT} vs. fixed point \eq{FP} vs. Weyl \eq{Weyl} consistency conditions (see Tab.~\ref{Tab}). The highest systematic perturbative approximation is NLO, or (222), where bounds in the range of $\eps_{\rm max}\approx 0.21$ arise through vacuum instability.  In the Weyl consistency scheme  2NLO${}^{\prime\prime}$, or (321), the bound is pushed towards $\eps_{\rm max}\approx 0.13$ due to mergers. In this work, we have argued that the consistent fixed point approximation  2NLO${}^{\prime}$, or (322), should be favoured. Its bound $\eps_{\rm max}\approx 0.09$ is even lower than the one in the Weyl scheme, and, as in the PT scheme, dominated by vacuum instability rather than mergers.  Taking the most advanced approximations as benchmarks, we conclude that the UV conformal window extends up to
\beq\label{epsbeta}
\eps_{\rm max}\approx 0.09\,\dots\, 0.13\,, 
\eeq 
see Fig.~\ref{pWindow}. The bounds \eq{epsbeta} from beta functions are stronger
than the bounds from their perturbative solutions \eq{max}. 
Also, all couplings and anomalous mass dimensions are still small (below $0.06$ and $0.15$, respectively) and in the range \eq{epsbeta} where perturbation theory is viable. 

In summary, competing effects due to higher loop contributions in the gauge,  scalar and Yukawa sector constrain the size of the UV conformal window.  While higher loop terms in the Yukawa sector continue to stabilise the fixed point, those in the gauge and scalar sector destabilise it. The combined effect is such that vacuum stability comes out as the most constraining factor.  Subleading terms in $\eps$ in all beta function coefficients always lead to tighter constraints \eq{compare}. The fact that the constraints for $\eps_{\rm subl.}$ and $\eps_{\rm strict}$ are quantitatively close to each other is a strong sign for the intrinsic consistency of results.

\subsection{Bounds from strong coupling}

We briefly comment on the prospect for asymptotic safety when $\eps$ becomes large \cite{Litim:2014uca}. Increasing $\eps$ implies that the one-loop term \eq{betag} is no longer small and perturbative control is lost. For an interacting fixed point to exist, cancellations between different loop orders must take place. For $N_F\to \infty$ and at finite $N_C$, corresponding to the limit $1/\eps\to 0$, the running of couplings is fully dominated by fermion loops, and gluon loops can be neglected. An infinite order resummation for the $U(1)$ \cite{PalanquesMestre:1983zy} and $SU(N)$ \cite{Gracey:1996he} beta functions can  be achieved, showing a non-perturbative UV fixed point in the gauge sector with $\eps\, \alpha_g^*$ of order unity \eq{couplings}. However, subleading corrections in $1/N_F$ may spoil the result and must be investigated before definite conclusions can be taken 
\cite{Holdom:2010qs}. Also, the Yukawa and scalar couplings do not play a role and can be omitted ($\alpha_y=\alpha_u=\al v=0$).  Based on continuity in $(N_F,N_C)$ it has been argued that a fingerprint of the fixed point should be visible at loop level \cite{Litim:2014uca}. Then, {\it assuming} that the UV fixed point exists non-perturbatively for sufficiently large and finite $N_F, N_C$, we may use the loop expansion to estimate a lower bound for its conformal window. Specifically, for large $\eps$, the leading $n$-loop contribution scales as $c_n\eps^{n-1} \alpha_g^{n+1}$ $(n>1)$ where $c_n$ is of order unity and independent of $\eps$. Cancellation with the one loop term gives the estimate $\alpha_g^*\sim \eps^{(2-n)/(n-1)}$ from the $n$th loop order ($c_n<0$) \cite{Pica:2010mt,Shrock:2013cca}.
Quantitatively, the three loop beta function \eq{betag} indicates that a strongly coupled fixed point obeys  $\eps>\eps_{\rm min}$,  with
   \beq\label{large}
   \eps_{\rm min}=\frac{3}{224} (159 + 19 \sqrt{505})\approx 7.49\,.
   \eeq
The bound arises from strong coupling with $\alpha_g\to\infty$ for $\eps\to \eps_{\rm min}$. Technically, it is due to a competition between subleading three loop terms and the two loop term. In the domain $\eps>\eps_{\rm min}$ the effective gauge coupling $\sqrt{\eps}\alpha^*_g$ is of order unity. The fixed point has one relevant eigendirection and the scaling exponent is large and bounded from above, $\vartheta(\eps)\le -20.69$, with $\eps\approx 44.6$ at the maximum. Moreover, 
the scaling exponent 
diverges $(\vartheta\to -\infty)$ at the bound \eq{large}, and in the limit $\eps\to \infty$ \cite{Litim:2014uca}. Hence, the expected characteristics of the fixed point at strong coupling are quite different from those at small $\eps$ where couplings and exponents are both parametrically small.

\subsection{Implications for model building and  cosmology}\label{pheno}
Finally, we discuss a few implications of our results for model building and cosmology \cite{Litim:2011cp}. It has already been shown that asymptotic safety offers novel opportunities for model building, including explicit BSM scenarios and phenomenological signatures with the Standard Model gauge group $SU(3)_C\times SU(2)_W\times U(1)_Y$ \cite{Bond:2017wut}. Moreover, estimates for the UV conformal window in terms of matter field multiplicities and representations have equally been derived \cite{Bond:2017wut}. 

For the model at hand, and using the bound \eq{max} from fixed points and scaling exponents at 2NLO${}'$, we obtain the smallest pair of integer values for $(N_C,N_F)$ compatible with asymptotic safety.  The first few integer solutions within \eq{max} are
\beq\label{maxN}
\begin{array}{rcl}
(N_C,N_F)&=&(3,17), (4,23), (5,28), (5,29), 
\\ &&
(6,34), (7,39), (7,40),\dots
\end{array}
\eeq
as indicated in the  yellow band of Fig.~\ref{pWindow1}.  Solutions cover all special unitary gauge groups with $N_C>2$. Starting from $N_C=5$ onwards, multiple solutions for the corresponding fermion flavour multiplicities $N_F$ become available.  
Bounds for the conformal window  from beta functions  \eq{epsbeta} are  tighter.
Considering the bound from (321), Tab.~\ref{tmax}, the first few integer solutions are
\beq\label{321}
\begin{array}{rcl}
(N_C,N_F)&=&
(5,28), (7,39), (8,45),(9,50),\\&&  (10,56), (11,61), (12,67),\dots
\end{array}
\eeq
corresponding to the entire yellow band in Fig.~\ref{pWindow}.  For the few leading values for $(N_C,N_F)$, the bound \eq{321} is the same irrespective of whether one uses the limit $\eps_{\rm subl.}$ (as has been done in \cite{Litim:2014uca}), or  the limit $\eps_{\rm strict}$. Moreover, solutions for $SU(3),\,SU(4)$ and $SU(6)$ are no longer available. The asymptotically safe solution with the smallest number of fields corresponds to  $SU(5)$ with $28$ flavours of fermions in the fundamental representation. 
This is quite close to the $SU(5)$ GUT candidate \cite{Georgi:1974sy}, which, with $N_F=24$ flavours of fermions, remains marginally asymptotically free.
Hence, \eq{321} suggests that asymptotic safety can already be achieved in a GUT-like scenario, with just a few more flavours of fermions (to destabilise asymptotic freedom), plus additional elementary mesons and Yukawa couplings (to generate asymptotic safety). Extending approximations to the complete (322) level, the bounds are shifted and the UV conformal window narrows down, starting with
\beq\label{322}
(N_C,N_F)=(7,39), (9,50), (11,61), (12,67),\dots
\eeq
corresponding to the lower yellow band in Fig.~\ref{pWindow}. Once again, the few leading integer solutions in \eq{322} do not depend on having used either $\eps_{\rm strict}$ or $\eps_{\rm subl.}$ to fix the conformal window. In particular, the cases $(N_C,N_F)=(5,28), (8,45)$ and  $(10,56)$ have dropped out due to the onset of vacuum instability in the fundamental meson sector, turning the first viable candidate into $SU(7)$. 

Asymptotic safety has also been considered as a mechanism for inflation by including UV effects from quantum gravity and matter \cite{Weinberg:2009wa,Hindmarsh:2011hx}. General scenarios have been classified and conditions for cosmological fixed points with inflationary expansions in the early universe are known  \cite{Hindmarsh:2011hx} (see  \cite{Falls:2016wsa,Falls:2017lst} for settings where inflation arises purely quantum gravitationally). It has also been speculated that inflation may arise from 
asymptotically safe toy models \cite{Litim:2014uca,Litim:2015iea}, neglecting quantum gravity  altogether \cite{Nielsen:2015una,Svendsen:2016kvn}. 
Compatibility with the 2015 Planck data \cite{Ade:2015lrj} at the 2$\sigma$-level requires a large conformal window up to $\eps\approx 0.7\,\dots\,0.8$ if minimal coupling is assumed \cite{Nielsen:2015una}. This scenario seems firmly excluded in the light of  \eq{max} and \eq{epsbeta}. Without minimal coupling, the conformal window \eq{epsbeta} imposes large values for the non-minimal coupling $\xi\gg 1$ of scalar matter to gravity (substantially larger than the conformal value $\xi=\016$) to achieve compatibility with data \cite{Ade:2015lrj}.

It is interesting to check how finite $N$ corrections beyond the Veneziano limit \cite{Medina:2016}, higher loop corrections beyond 2NLO${}^\prime$, higher-dimensional operators \cite{Buyukbese:2017ehm}, or strong coupling effects, are going to modify  the  UV conformal window and the bounds \eq{epsbeta}, \eq{321} and \eq{322}. This is left for future work.

\section{\bf Discussion}\label{discussion}
The existence of exact and interacting UV fixed points in particle physics offers many opportunities for model building \cite{Bond:2017wut}. For any practical applications, however, it is equally important to  understand the size of the corresponding conformal window. Here, we have investigated the conformal window for the gauge-Yukawa theory \eq{L}. Extending the findings of \cite{Litim:2014uca}
we have obtained exact results for fixed points, anomalous dimensions, and scaling exponents up to second order in the small parameter \eq{eps},
the highest order in perturbation theory presently  available. 
The underlying ordering principle, which due to the fixed point is different from what one would expect normally, is  also explained in detail (Tab.~\ref{Tab}).

The conformal window follows from fixed points and beta functions.
We  have also compared different approximation orders  and clarified the role of subleading corrections (Tab.~\ref{TabEps}).  
Limits invariably arise through a competition of fluctuations. Higher loops in the Yukawa sector enhance the conformal window, countered by higher loops in the gauge sector. Higher loops in the scalar sector tend to destabilise the quantum vacuum. With increasing coupling strength, the conformal window terminates either through fixed point mergers or via the onset of vacuum instability.
Despite their qualitatively different origins, constraints  are quantitatively similar, with vacuum stability offering the tightest one, \eq{epsbeta}. 
Moreover, the conformal window  based on  the convergence of fixed points and scaling exponents (Fig.~\ref{pWindow1}) is less constrained than the one based on beta functions (Fig.~\ref{pWindow}). 
Some phenomenological implications have been worked out for particle physics and cosmology.

It has also been noted that another conformal  window may exist in the regime where the parameter $\eps$ becomes large \cite{PalanquesMestre:1983zy,Gracey:1996he,Holdom:2010qs,Litim:2014uca}. If so, the underlying mechanism is non-perturbative. Presently, results are available at the leading order in $1/\eps$. Assuming the fixed point exists at finite $1/\eps$, a rough estimate for its conformal window has been given based on perturbation theory, \eq{large}. 

As a final point, we note that the theory remains perturbative in the entire conformal window, much unlike the IR conformal windows in QCD-like theories \cite{DelDebbio:2010zz}. The culprit for this is the scalar sector which controls the stability of the ground state.
It would be good to confirm  these results non-perturbatively, also in view of higher dimensional operators and finite $N$ corrections. 

\section*{\bf Acknowledgements}

This work is supported by the {\it Deutsche Akademische Austauschdienst} (DAAD) under the  Grant [57314657], by the {\it Consejo Nacional de Ciencia y Tecnolog\'ia} (CONACYT),   and by a studentship from the {\it Science and Technology Facilities Council} (STFC).

\appendix

\section{Technicalities}
\label{Tech}
In Sec.~\ref{AS}, and starting from known general expressions  in the $\overline{\mathrm{MS}}$ renormalisation scheme 
\cite{Machacek:1983tz,Machacek:1983fi,Machacek:1984zw,Luo:2002ti,Pickering:2001aq}, we have derived all beta functions and anomalous dimensions for our model both manually, and with the help of a purpose-made algebraic code.  In this appendix we provide some details on the  extraction of the two-loop contributions to the running of the scalar quartic couplings. We follow closely the notation of \cite{Luo:2002ti} and \cite{Machacek:1983tz, Machacek:1983fi, Machacek:1984zw}. Our conventions for the most general Yukawa and quartic scalar selfinteractions  are
\begin{equation}\label{Yukpot}
 \begin{array}{rcl}
  L_{\rm Yuk.}&=&-\tfrac{1}{2}(Y^a_{jk}\, \Phi^a\Psi_j\Psi_k + {\rm h.c.})\,,\\[1ex]
  L_{\rm pot.}&=& -\frac{1}{4!}\lambda_{abcd}\,\Phi^a\Phi^b\Phi^c\Phi^d\,,
  \end{array}
  \end{equation}
  where $\Psi_j$ denote Weyl fermions, and $\Phi^a$ real scalars.   Below, we will find it  convenient to view the Yukawa couplings as symmetric matrices  in the fermionic indices $Y^a$, with  $(Y^a)_{jk}=Y^a_{jk}$.

Due to the scalars being gauge singlets in our model \eq{L}, \eq{Lall},  the number of non-zero contributions reduces drastically, and a general expression for the two-loop beta function of the quartics can be given. Writing the scalar beta functions as $\beta_{abcd}\equiv \mu\partial_\mu \lambda_{abcd}$, and also using  conventions as in \eq{betaPT}, we have
\begin{widetext}
\begin{align}
\beta^{(2)}_{abcd}
= & \sum_{e=a,b,c,d} \frac{1}{2} \left( \Lambda^2_{ee} - 3 H^2_{ee} - 2 \overline{H}^2_{ee} + 10 Y^{2F}_{ee}\right) \lambda_{abcd}  \nonumber \\
&- \overline{\Lambda}^{3}_{abcd} - 2 \overline{\Lambda}^{2Y}_{abcd} + \overline{H}^\lambda_{abcd} + 2 H^Y_{abcd} + 4 \overline{H}^Y_{abcd} + 4 H^3_{abcd} - 2 H^F_{abcd}\,. \label{QuarticBetaParameter}
\end{align} 
\end{widetext}
For convenience, we have  scaled the loop factor $(4\pi)^4$ into the 
couplings. The terms in the first line of \eq{QuarticBetaParameter} are the two-loop corrections to the scalar legs, with
\beq\label{legs}
\begin{array}{rl}
\Lambda^2_{ab} &= 
\tfrac{1}{6} \lambda_{acde} \lambda_{bcde}, 
\\[1ex]
H^2_{ab} &= 
\tfrac{1}{2} \mathrm{Tr}\left[ Y^a Y^{\dagger b} Y^c Y^{\dagger c} + Y^{\dagger a} Y^{b} Y^{\dagger c} Y^{c} \right], \\[1ex]
\overline{H}^2_{ab} &= 
\tfrac{1}{2} \mathrm{Tr}\left[ Y^a Y^{\dagger c} Y^b Y^{\dagger c} + Y^{\dagger a} Y^{c} Y^{\dagger b} Y^{c} \right], \\[1ex]
Y^{2F}_{ab} &= \tfrac{1}{2} \,g^2\, \mathrm{Tr}\left[ C_2(F) \left(Y^a Y^{\dagger b} + Y^{b} Y^{\dagger a}\right)\right]\,.
\end{array}
\eeq
The terms in the second line of \eq{QuarticBetaParameter}  are the  various vertex corrections, defined as
\begin{align}
\overline{\Lambda}^3_{abcd} &= \tfrac{1}{4} \sum_\mathrm{perms} \lambda_{abef} \lambda_{cegh} \lambda_{dfgh}, 
\nonumber\\
\overline{\Lambda}^{2Y}_{abcd} &= \tfrac{1}{16} \sum_\mathrm{perms} \lambda_{abef} \lambda_{cdeg} \mathrm{Tr}\left[ Y^{\dagger f} Y^g + Y^{\dagger g} Y^f\right],\nonumber\\
\overline{H}^\lambda_{abcd} &= \tfrac{1}{8} \sum_\mathrm{perms} \lambda_{abef} \mathrm{Tr}\left[ Y^{c} Y^{\dagger e} Y^{d} Y^{\dagger f} +(Y\leftrightarrow Y^\dagger)
\right], \nonumber\\
H^Y_{abcd} &= \sum_\mathrm{perms} \mathrm{Tr}\left[ Y^{ \dagger a} Y^{ b} Y^{\dagger c} Y^{ d} Y^{\dagger e} Y^{ e} \right], \label{vertex}
\\
\overline{H}^Y_{abcd} &= \tfrac{1}{2} \sum_\mathrm{perms} \mathrm{Tr}\left[ Y^{ \dagger a} Y^{ e} Y^{\dagger b} Y^{ c} Y^{\dagger d} Y^{e} 
+(Y\leftrightarrow Y^\dagger)
\nonumber
\right]
\\
H^3_{abcd} &= \tfrac{1}{2} \sum_\mathrm{perms} \mathrm{Tr}\left[Y^{ a}  Y^{ \dagger b} Y^{ e}  Y^{\dagger c} Y^{ d} Y^{\dagger e} \right],\nonumber \\
H^F_{abcd} &= g^2 \sum_\mathrm{perms} \mathrm{Tr}\left[\{C_2(F), Y^a\} Y^{\dagger b} Y^c Y^{\dagger d}  \right],\nonumber
\end{align}
where $\sum_\mathrm{perms}$ denotes the sum over all permutations of the indices $a,\,b,\,c,\,d$. Traces are taken over all fermion indices, and the matrix $C_2(F)$ is the quadratic Casimir for the fermions.

Next, we need to map and evaluate expressions in the conventions of our model \eq{L}, \eq{Lall} and  \eq{couplings}. The algebra is somewhat tedious since the scalar couplings $\lambda_{abcd}$ in \eq{Yukpot} are fully symmetrised, differently normalised than those in the model considered here, and defined in terms of fields decomposed into real degrees of freedom. One simplification is that the contribution from the field strength renormalisation is, of course, equal for each of the quartic couplings $\lambda_{abcd}$. By a suitable choice of outer indices, renormalisation group equations  for $\alpha_u,\,\alpha_v$ in \eq{couplings} are obtained.
{For example, for the double-trace coupling, taking the outer legs as $\Phi^a,\Phi^b = ({\rm Re}\, H)_{i i}$ and $ \Phi^c,\Phi^d = ({\rm Re}\, H)_{jj}$ with $i \neq j$, leads to  $\frac{1}{4!} \lambda_{aacc} = \alpha_v/(12 N_F^2)$. For the single trace coupling, taking $\Phi^a = ({\rm Re}\, H)_{i i} , \Phi^b = ({\rm Re}\,  H)_{i j}, \Phi^c = ({\rm Re}\,  H)_{j j}$ and  $\Phi^d = ({\rm Re}\,  H)_{ji}$ with $i \neq j $ leads to $\frac{1}{4!} \lambda_{abcd} =  \alpha_u/(24 N_F)$, and similarily for the map from $Y^a_{jk}$ onto $\alpha_y$.} 

With these considerations in mind we find the two loop contributions to $\mu\partial_\mu 
\alpha_{u,v}$ from \eqref{QuarticBetaParameter}, \eq{legs} and \eq{vertex}. In terms of \eq{eps}, 
  and neglecting subleading terms of ${\cal O}(1/N)$ in the Veneziano limit,
we obtain from \eq{legs}
\beq\label{legsuv}
\begin{array}{rl}
\sum_e \Lambda^2_{ee} &= 16 \alpha_u^2\,, 
\\[1ex]
\sum_e H^2_{ee} &= 2 (11 + 2 \epsilon) \alpha_y^2\,,  \\[1ex]
\sum_e \overline{H}^2_{ee}  &= 0\,, \\[1ex]
\sum_e Y^{2F}_{ee} &= 2 \alpha_g \alpha_y\,,
\end{array}
\eeq
where the sum runs over any four scalar indices. The two-loop vertex corrections   \eq{vertex}  to the flow of the single-trace quartic coupling $\mu\partial_\mu\alpha_u$, normalised to account for the map from $\lambda_{abcd}$ to $\alpha_u$ in \eqref{QuarticBetaParameter}, are 
\begin{align}
\overline{\Lambda}^3_{u}
 &= 32 \alpha_u^3 \,,\nonumber\\
\overline{\Lambda}^{2Y}_{u}
&= 8 \alpha_y \alpha_u^2\,,\nonumber\\
\overline{H}^\lambda_{u}
&= 0 \,,\nonumber\\
H^Y_{u}
&=  \tfrac{1}{2}\left(11 + 2 \epsilon\right)^2 \alpha_y^3 \,, \label{vertexv}\\
\overline{H}^Y_{u}
&=  0\,,\nonumber\\
H^3_{u} 
&= 0\,,\nonumber\\
H^F_{u} 
&= (11 + 2\epsilon) \alpha_g \alpha_y^2\,.\nonumber
\end{align}
Similarly, the vertex corrections  \eq{vertex} to the flow of the double-trace coupling $\mu\partial_\mu \alpha_{v}$,
now normalised to account for the map from $\lambda_{abcd}$ to $\alpha_v$, are given by
\begin{align}
\overline{\Lambda}^3_{v} 
&= 48\,\alpha_u^2\,(2 \alpha_u  + \alpha_v)\,,\nonumber\\
\overline{\Lambda}^{2Y}_{v}
&= 4\,\alpha_y \left(3 \alpha_u^2 + 4\alpha_u \alpha_v + \alpha_v^2\right)\,,
\nonumber\\
\overline{H}^\lambda_{v}
&= 4(11 + 2 \epsilon) \alpha_y^2 \alpha_u \,,\nonumber\\
H^Y_{v}
&=  0 \label{vertexu}\,, 
\\
\overline{H}^Y_{v}
&=  0\,,\nonumber\\
H^3_{v} 
&= \tfrac{1}{4} (11 + 2\epsilon)^2 \alpha_y^3\,,\nonumber\\
H^F_{v}
&= 0\nonumber \,.
\end{align}
Combining 
\eq{legsuv}, \eq{vertexv} and \eq{vertexu}  
leads to  the  final result  \eqref{betau} and \eqref{betav}.
The expressions for the two-loop anomalous dimensions \eq{gamma}, \eq{gammaM} have been deduced from general expressions using  similar techniques.

\bibliographystyle{JHEP_withtitle}
\bibliography{bib_DFL}

\end{document}